\documentclass{article}
\usepackage{amssymb}

\begin{document}

\begin{center}
{\Large \bf On calculating the mean values of quantum observables in the optical tomography representation}
\end{center}

\begin{center} {\bf G. G. Amosov$^1$, Ya. A. Korennoy$^2$, V. I. Man'ko$^2$ }\end{center}

\smallskip

\begin{center}
{\it $^1$Steklov Mathematical Institute \\
ul. Gubkina 8, Moscow 119991, Russia}\\
{\it $^2$P.N.    Lebedev Physics Institute,                          \\
       Leninsky prospect 53, 117924 Moscow, Russia }
\end{center}

\begin {abstract}

Given a density operator $\hat \rho$ the optical tomography map defines a one-parameter set of probability distributions
$w_{\hat \rho}(X,\phi),\ \phi \in [0,2\pi ),$ on the real line  allowing to reconstruct $\hat \rho $. We introduce a dual map from
the special class $\mathcal A$ of quantum observables $\hat a$ to a special class of generalized functions $a(X,\phi)$ such that
the mean value $<\hat a>_{\hat \rho} =Tr(\hat \rho\hat a)$ is given by the formula $<\hat a>_{\hat \rho}=
\int \limits _{0}^{2\pi}\int \limits _{-\infty}^{+\infty }w_{\hat \rho}(X,\phi)a(X,\phi)dXd\phi$. The class $\mathcal A$ includes
all the symmetrized polynomials of canonical variables $\hat q$ and $\hat p$.

\end{abstract}

\section {Introduction}

Given an observable (hermitian operator) $\hat a$ in a Hilbert space $H$ the spectral theorem reads
$$
\hat a=\int \limits _{\mathbb R}Xd\hat E((-\infty ,X]),
$$
where $\hat E$ in an orthogonal projection valued measure defined on all
Borel subsets $\Omega \subset {\mathbb R}$ such that $\hat E(\Omega )$ is
an orthogonal projection and the projections $\hat E(\Omega _1), \ \hat E(\Omega _2)$
are orthogonal for all open $\Omega _1,\Omega _2\subset {\mathbb R},\ \Omega _1\cap \Omega _2=\emptyset$.
Using the projection valued (spectral) measure $\hat E$ transforms the Hilbert space $H$ to the Hilbert space
$H_{\hat a}=L^2({\mathbb R})$ formed by wave functions $\psi _{\hat a}(\cdot)$ obtaining from $\psi \in H$
by the formula
$$
\psi_{\hat a} (X)=\frac {d}{dX}\left (\hat E((-\infty ,X])\psi \right ).
$$
The Hilbert space $H_{\hat a}$ is said to be a space of representation associated with the observable $\hat a$.

Suppose that $\hat \rho $ is a density  operator (positive unit-trace operator), then in any space of representation
$H_{\hat a}$ it can be represented as an integral operator
$$
(\hat \rho \psi _{\hat a})(X)=\int \limits _{\mathbb R}\rho _{\hat a}(X,Y)\psi_{\hat a} (Y)dY,
$$
$\psi _{\hat a}(\cdot )\in H_{\hat a}$. In the case, the Hilbert-Schmidt kernel $\rho _{\hat a}(\cdot ,\cdot)$ is said to be a density matrix of $\hat \rho$
in the space of representation $H_{\hat a}$. Analogously, one can define the density matrix $b(\cdot,\cdot)$ (which can be a
generalized function) associated with a observable $\hat b$ in the space of representation $H_{\hat a}$.

In \cite {Wig32} the Wigner function $W(q,p)$ associated with the density matrix
$\hat \rho (\cdot ,\cdot)$ in the space of representation associated with the position operator $\hat q$ was introduced as
$$
W(q,p)=\frac {1}{2\pi}\int \limits _{\mathbb R}e^{-ipx}\rho \left (q+\frac {x}{2},q-\frac {x}{2}\right )dx.
$$
The Moyal representation of quantum mechanics \cite {Moy49} defines a map between quantum observables $\hat a$
and functions $a(q,p)$ on the phase space under which the mean value $<\hat a>_{\hat \rho}=Tr(\hat \rho\hat a)$ is
given by the formula
$$
<\hat a>_{\hat \rho}=\int \limits _{-\infty}^{+\infty}\int \limits _{-\infty}^{+\infty}W(q,p)a(q,p)dqdp.
$$

Unfortunately, although the normalization rule $\int \int W(q,p)dqdp=1$ holds, the Wigner function $W(q,p)$ is not positive definite in general.
In \cite {Ber87, Vog89} the optical tomogram $w(X,\phi)$ which can be calculated under experimental measuring a generalized homodyne quadrature
was introduced as the Radon transform of the Wigner function,
$$
w(X,\phi)=\int \int W(q,p)\delta (X-\cos (\phi )q-\sin (\phi)p)dqdp,
$$
where $\hat q$ and $\hat p$ are the position and momentum operators. The one-parameter set $\{w(X,\phi),\ \phi \in [0,2\pi)\}$ consists of
probability distributions on the real line. The optical tomogram can be calculated from the density operator directly by means of
the formula \cite {Man95}
$$
w(X,\phi)=Tr(\hat \rho \delta (X-\cos(\phi)\hat q-\sin(\phi)\hat p)).
$$
The inverse Radon transform \cite{dAr95} allows to reconstruct the Wigner function from the optical tomogram.

For a density operator $\hat a$ one can define a function of complex variable $z$ by the formula
\begin{equation}\label{dual}
a(z,\phi)=-2\pi Tr(\hat a (z-\cos (\phi )\hat q-\sin(\phi )\hat p)^{-2}),
\end{equation}
$z\in {\mathbb C},\ Im(z)\neq 0,\ \phi \in [0,2\pi]$.

In the present paper we shall correct the mistake in \cite {AM09}. Our goal is to prove the following statements.

{\bf Theorem 1.} {\it  For any density operator
$\hat \rho$ the following identity holds,
$$
\lim \limits _{\varepsilon \to +0}\int \limits _{0}^{2\pi}\int \limits _{-\infty }^{+\infty}w_{\hat \rho}(X+i\varepsilon,\phi)a(X+i\varepsilon,\phi)dXd\phi=Tr(\hat \rho\hat a).
$$
}

{\bf Definition.} We shall call the relation (\ref {dual}) {\it a map dual to the optical tomogram map}.

It should be noted that the notion of duality we introduce is different from the known concept of \cite{OM97}.

Denote $\mathcal D$ the convex set of density operators whose kernels in the coordinate representation belong to
the Schwartz space $S({\mathbb R}^2)$. Then, optical tomograms corresponding to states from $\mathcal D$
belong to the space $\Omega $ consisting of functions $w(X,\phi)$ which are from the Schwartz space in $x$
and infinitely differentiable in $\phi $.
Notice that ${\mathcal A}={\mathcal D}^*$ contains all bounded quantum
observables at least.

{\bf Corollary 2.} {\it The dual map (\ref {dual}) can be extended to
any $\hat a\in {\mathcal A}$. The extension $a(X,\phi)$ belongs to the adjoint space $\Omega ^*$.
Moreover, for any density
operator $\hat \rho \in {\mathcal D}$ the equality
$$
\int \limits _{0}^{2\pi}\int \limits _{-\infty}^{+\infty}w_{\hat \rho}(X,\phi)a(X,\phi)dXd\phi=
Tr(\hat \rho \hat a)
$$
holds.

}

Let us define a symmetrized product of canonical quantum observables $\hat q^m\hat p^n$ as
\begin{equation}\label{sym}
\{\hat q^m\hat p^n\}_s=\frac {1}{2^n}\sum \limits _{k=0}^nC_{n}^{k}\hat p^{k}\hat q^m\hat p^{n-k}.
\end{equation}

Below we use the trigonometric polynomials $Q_n^m(\cos(\phi))$ defined in Appendix.

{\bf Theorem 3.}{\it The action of the dual map (\ref {dual}) to the observables (\ref {sym}),
gives rise to $a_{mn}(X,\phi)$ of the form
$$
a_{mn}(X,\phi)=Q_{n+m}^m(\cos(\phi))X^{n+m}.
$$
}

\section {The Parseval equality associated with the characteristic functions}

Given a density operator $\hat \rho$ the function $F(\mu ,\nu)=Tr(\hat \rho e^{i\mu \hat q+i\nu \hat p})$ is said to be a characteristic
function of the state $\hat \rho$. The associated set of probability distributions is said to be {\it a symplectic quantum tomogram}
\cite {Man95}
$$
w(X,\mu ,\nu)=\frac {1}{2\pi}\int \limits _{-\infty}^{+\infty}e^{-iXt}F(t\mu,t\nu)dt
$$
which is connected with the optical tomogram by the formula
$$
w(X,\phi)=w(X,\cos(\phi),\sin(\phi)).
$$
In this way,
\begin{equation}\label{con}
F(t\cos(\phi) ,t\sin(\phi))=\int \limits _{-\infty}^{+\infty}e^{itX}w(X,\phi)dX.
\end{equation}

The standard identity $e^{i\mu \hat q+i\nu \hat p}=e^{\frac {i\mu \nu}{2}}e^{i\mu \hat q}e^{i\nu \hat p}$
results in
\begin{equation}\label{char}
F(\mu ,\nu)=\int \limits _{-\infty}^{+\infty}e^{i\mu x}\rho \left (x+\frac {\nu}{2},x-\frac {\nu}{2}\right )dx.
\end{equation}
It immediately follows from (\ref {char}) that the following Parseval-type equality holds,
$$
\int \limits _{-\infty }^{+\infty}\int \limits _{-\infty}^{+\infty}|F(\mu ,\nu)|^2d\mu\nu=\frac {1}{2\pi}\int \limits _{-\infty }^{+\infty}\int \limits _{-\infty}^{+\infty}\rho (X,Y)dXdY=
\frac{1}{2\pi}Tr(\hat \rho ^2),
$$
which is equivalent to
\begin{equation}\label{par}
\int \limits _{-\infty }^{+\infty}\int \limits _{-\infty}^{+\infty}F_{\hat \rho}(\mu ,\nu)\overline F_{\hat \sigma}(\mu ,\nu)d\mu\nu=
\frac {1}{2\pi}Tr(\hat \rho \hat \sigma)
\end{equation}
for the characteristic functions of any two density operators $\hat \rho $ and $\hat \sigma$.

Taking into account the Parseval-type equality (\ref {par}) it is possible to extend the map $\hat \rho \to F_{\hat \rho }$
to all operators of Hilbert-Schmidt class. Moreover, one can construct a tempered distribution $F_{\hat a}\in S'({\mathbb R}^2)$
associated with an observable $\hat a$ such that $\int \limits _{-\infty}^{+\infty}\int \limits _{-\infty}^{+\infty}
F_{\hat \rho}(\mu,\nu)F_{\hat a}(\mu ,\nu)d\mu d\nu=\frac {1}{2\pi}Tr(\hat \rho\hat a)$ for all density operators $\hat \rho\in {\mathcal D}$.
The following result is well-known (\cite {Moy49}) and we put it for the sake of completeness.

{\bf Proposition 4.}{\it The tempered distributions $F_{\hat a}\equiv F_{mn}$ associated with the observables $\hat a$ of
the form (\ref {sym}) are given by the formula
$$
F_{mn}(\mu ,\nu)=(-i)^{m+n}\delta ^{(m)}(\mu )\delta ^{(n)}(\nu).
$$
}

Proof.

Using the Parseval type identity (\ref {par}) we get
$$
\int \limits _{-\infty}^{+\infty}F_{\hat\rho}(\mu ,\nu)F_{00}(\mu ,\nu)d\mu d\nu =\frac {1}{2\pi}Tr(\hat\rho)=\frac {1}{2\pi}.
$$
Since $F_{\hat \rho}(0,0)=1$ for all density operators $\hat \rho$ it results in
\begin{equation}\label{id}
F_{00}(\mu ,\nu)=\delta (\mu )\delta (\nu).
\end{equation}
Notice that the statement holds if either $m$ or $n$ equals zero. Suppose that it is true for all integer numbers up to fixed $m$ and $n$, let us prove
that it holds for $m+1$ and $n+1$. Using the equalities
$$
\hat p^k\hat q^m\hat p^{n-k}=\hat p^k\hat q^{m+1}\hat p^{n-k}-i(n-k)\hat p^k\hat q^m\hat p^{n-k-1}
$$
and
$$
\nu\delta (\nu)=0,\ \nu \delta ^{(n)}(\nu)=-n\delta ^{(n-1)}(\nu),\ n\ge 1,
$$
we get
$$
\frac {\partial }{\partial \mu}\left (Tr\left (\{\hat q^m\hat p^n\}_se^{i\mu \hat q+i\nu \hat p}\right )\right )=Tr\left (\{\hat q^m\hat p^n\}_s(i\hat q+\frac {i\nu}{2})e^{\frac {i\mu\nu}{2}}e^{i\mu \hat q}e^{i\nu \hat p}\right )=
$$
$$
iTr\left (\{\hat q^m\hat p^n\}_s\hat qe^{i\mu \hat q+i\nu \hat p}\right )-\frac {i}{2}n\delta ^{(m)}(\mu)\delta ^{(n-1)}(\nu)=iTr\left (\{\hat q^{m+1}\hat p^n\}_se^{i\mu \hat q+i\nu \hat p}\right ).
$$
On the other hand, the equality
$$
\frac {\hat p}{2}\{\hat q^m\hat p^n\}_s+\{\hat q^m\hat p^n\}_s\frac {\hat p}{2}=\{\hat q^m\hat p^{n+1}\}_s
$$
results in
$$
\frac {\partial}{\partial \nu}\left (Tr\left (\{\hat q^m\hat p^n\}_se^{i\mu \hat q+i\nu \hat p}\right )\right )=Tr(\{\hat q^m\hat p^n\}_se^{\frac {i\mu\nu}{2}}e^{i\mu \hat q}(i\hat p+\frac {i\mu}{2})e^{i\nu \hat p})=
$$
$$
Tr(\frac {ip}{2}\{\hat q^m\hat p^n\}_se^{\frac {i\mu\nu}{2}}e^{i\mu \hat q}e^{i\nu \hat p})+Tr(\{\hat q^m\hat p^n\}_se^{\frac {i\mu\nu}{2}}(\frac {i\hat p}{2}-\frac {i\mu}{2})e^{i\mu \hat q}e^{i\nu \hat p})+\frac {im}{2}\delta ^{(m-1)}(\mu)\delta ^{(n)}(\nu)=
$$
$$
iTr(\{\hat q^{m}\hat p^{n+1}\}_se^{i\mu\hat q+i\nu \hat p}).
$$
$\Box$

\section{The dual map}

To prove Theorem 1 and Corollary 2 we need the following result.

{\bf Proposition 5.}{\it Given a density operator $\hat a$ the relation between the dual map  (\ref {dual}) and the characteristic function $F_{\hat a}$ is
given by
$$
tF_{\hat a}(t\cos(\phi) ,t\sin(\phi))=\frac {1}{(2\pi )^2}\lim \limits _{\varepsilon \to +0}\int \limits _{-\infty}^{+\infty}e^{itX}a(X-i\varepsilon,\phi)dX,\ t>0.
$$
}
Proof.

Let us consider the representation of $\cos (\phi)\hat p+\sin(\phi)\hat q$ in the space $H_{\phi}=L^2({\mathbb R})$ such that
$$
((\cos (\phi)\hat p+\sin(\phi)\hat q)f)(x)=xf(x),\ f\in H_{\phi}.
$$
Then, given $f,g\in H_{\phi}$
$$
\int \limits _{-\infty}^{+\infty}e^{itX}(g,(X-i\varepsilon-\cos (\phi)\hat p-\sin(\phi)\hat q)^{-2}f)dX=
$$
$$
\int \limits _{-\infty}^{+\infty}\overline  g(x)f(x)\int \limits _{-\infty}^{+\infty} e^{itX}\frac {1}{(X-x-i\varepsilon )^2}dXdx\equiv I
$$

Calculating the residue in $z_0=x+i\varepsilon$ we obtain
$$
I=2\pi i\left \{\begin{array}{c}it(g,e^{it(\cos(\phi)\hat q+\sin(\phi)\hat p+i\varepsilon)}f),\ t>0\\ 0,\ t<0\end{array}
\right .
$$

$\Box$

Proof of Theorem 1.

Using the expression of $w_{\hat \rho}$ through the characteristic function $F_{\hat \rho}$ and the definition of $a(z,\phi)$ we obtain
$$
\lim \limits _{\epsilon \to +0}\int \limits _{0}^{2\pi}\int \limits _{-\infty }^{+\infty}w_{\hat \rho}(X+i\varepsilon,\phi)a(X+i\varepsilon,\phi)dXd\phi=
$$
$$
-\lim \limits _{\epsilon \to +0}\int \limits _{0}^{2\pi}\int \limits _{-\infty }^{+\infty}\int \limits _{-\infty }^{+\infty}
e^{-itX}F_{\hat \rho}(t\cos(\phi),t\sin(\phi))Tr(\hat a (X+i\varepsilon-\cos(\phi)\hat q-\sin(\phi)\hat p)^{-2})dtdXd\phi=
$$
$$
-\lim \limits _{\epsilon \to +0}\int \limits _{0}^{2\pi}\int \limits _{-\infty }^{+\infty}
F_{\hat \rho}(t\cos(\phi),t\sin(\phi))\left (\int \limits _{-\infty }^{+\infty}Tr(\hat a e^{-itX}(X+i\varepsilon -\cos(\phi)\hat q-\sin(\phi)\hat p)^{-2})dX\right )dtd\phi=
$$
$$
2\pi \lim \limits _{\epsilon \to 0}\int \limits _{0}^{2\pi}\int \limits _{-\infty }^{+\infty}
F_{\hat \rho}(t\cos(\phi),t\sin(\phi)\left (\overline {-\frac {1}{2\pi}\int \limits _{-\infty }^{+\infty}Tr(\hat a e^{itX}(X-i\varepsilon -\cos(\phi)\hat q-\sin(\phi)\hat p)^{-2}dX}\right )dtd\phi\equiv I
$$
Substituting the relation of Proposition 5
we get
$$
I=2\pi \int \limits _{-\infty}^{+\infty}\int\limits _{-\infty}^{+\infty}F_{\hat \rho}(\mu ,\nu)\overline F_{\hat a}(\mu ,\nu)d\mu d\nu=Tr(\hat \rho\hat a)
$$

$\Box$

Proof of Corollary 2.

If a density operator $\hat \rho \in {\mathcal D}$, i.e. the density matrix in the coordinate representation
$\rho (\cdot ,\cdot)\in S({\mathbb R}^2)$, then its characteristic function
$$
F_{\hat \rho}(\mu ,\nu)=\int \limits _{-\infty}^{+\infty}e^{i\mu x}\rho \left (x+\frac {\nu}{2},x-\frac {\nu}{2}\right )dx
$$
is also from the Schwartz space $S({\mathbb R}^2)$. Thus, the corresponding
optical tomogram
$$
\omega (X,\phi)=\frac {1}{2\pi}\int \limits _{-\infty}^{+\infty}e^{-itX}F_{\hat \rho}(\cos(\phi)t,\sin(\phi)t)dt
$$
belongs to $S({\mathbb R})$ in $X$ and infinitely differentiable in $\phi$.
Using the Parseval type equation of Theorem 1
$$
\lim \limits _{\varepsilon \to +0}\int \limits _{0}^{2\pi}\int \limits _{-\infty }^{+\infty}w_{\hat \rho}(X+i\varepsilon,\phi)a(X+i\varepsilon,\phi)dXd\phi=Tr(\hat \rho\hat a)
$$
we can define the extension of dual tomographic map $\hat a\to a(X,\phi)$ such that $a(X,\phi)$ should be a generalized function on the set $\Omega $ of
optical tomograms $\omega _{\hat \rho}$ such that
$$
<a,\omega_{\hat \rho}>=Tr(\hat \rho\hat a).
$$

$\Box $

Proof of Theorem 3.

Given an optical tomogram $\omega _{\hat \rho}(X,\phi)$ of a density operator $\hat \rho$ we get
$$
\int \limits _{-\infty}^{+\infty}\int \limits _{0}^{2\pi}X^{n+m}Q_{n+m}^m(\cos(\phi))\omega_{\hat \rho}(X,\phi)dXd\phi =
$$
$$
\frac {1}{2\pi}\int \limits _{-\infty}^{+\infty}\int \limits _0^{2\pi}X^{n+m}Q_{n+m}^m(\cos(\phi))\int \limits _{-\infty}
^{+\infty}e^{-iXt}F_{\hat \rho}(t\cos(\phi),t\sin(\phi))dtdXd\phi =
$$
$$
i^n\int \limits _{-\infty}^{+\infty}\int \limits _0^{2\pi}\delta ^{(n+m)}(t)F_{\hat \rho}(t\cos \phi,t\sin(\phi))Q_{n+m}^m(\cos(\phi))dtd\phi=
$$
$$
(-i)^n\int \limits _0^{2\pi}\sum \limits _{k=0}^{n+m}C_{n+m}^k\frac {\partial ^{n+m}F_{\hat \rho}}{\partial \mu ^k\partial \nu ^{n+m-k}}(0,0)
\cos ^k(\phi)\sin^{m+n-k}(\phi)Q_{n+m}^m(\cos(\phi))d\phi=
$$
$$
(-i)^n\frac {\partial ^{n+m}F_{\hat \rho}}{\partial \mu ^m\partial \nu ^{n}}(0,0).
$$
Now the result follows from Proposition 4. $\Box$

\section*{Appendix}

Let us consider the trigonometric system $\{\sin^k(\phi)\cos^{n-k}(\phi),\ 0\le k\le n\}$.
Taking derivatives of $\sin^k(\phi)\cos^{n-k}(\phi)$ give rise to linear combinations of
these elements. It follows that $\sin^k(\phi)\cos^{n-k}(\phi)$ satisfy to the linear differential
equation of $n+1$th order.
Notice that
$$
(\sin ^k(\phi)\cos ^{n-k}(\phi))^{(s)}=0,\ 0\le s<k,\ (\sin ^k(\phi)\cos ^{n-k}(\phi))^{(k)}=k!,\ if\ \phi =0.
$$
Hence the Wronskian $w(0)=\prod \limits _{k=0}^nk!\neq 0$ and  the elements
of this system are linear independent on the segment $[0,2\pi]$.
Thus, there exists the biorthogonal system $\tilde Q_n^m(\cos(\phi))$ consisting of polynomials in
$\sin ^k(\phi)\cos ^{n-k}(\phi)$ such that
$$
\int \limits _0^{2\pi}\sin ^k(\phi)\cos ^{n-k}(\phi)\tilde Q_n^m(\cos(\phi))d\phi=\delta _{km}.
$$
Put $Q_n^m(\cos (\phi))=\frac {1}{C_n^m}\tilde Q_n^m(\cos(\phi))$. The first several polynomials are
$$
Q_0^0(\cos(\phi))=\frac {1}{2\pi},\ Q_1^0(\cos(\phi))=\frac {1}{\pi}\cos(\phi),\ Q_1^1(\cos(\phi))=\frac {1}{\pi}\sin(\phi),
$$
$$
Q_2^0(\cos(\phi))=-\frac {1}{2\pi}\cos^2(\phi)+\frac {3}{2\pi}\sin^2(\phi),
$$
$$
Q_2^1(\cos(\phi))=\frac {2}{\pi}\sin(\phi)\cos(\phi),
$$
$$
Q_2^2(\cos(\phi))=\frac {3}{2\pi}\cos^2(\phi)-\frac{1}{2\pi}\sin^2(\phi).
$$

\section*{Acknowledgments} The work of GGA and VIM is partially supported by RFBR, grant 09-02-00142, 10-02-00312, 11-02-00456.

\end{document}